\documentclass[letterpaper]{article} 
\usepackage{aaai25}  
\usepackage{times}  
\usepackage{helvet}  
\usepackage{courier}  
\usepackage[hyphens]{url}  
\usepackage{graphicx} 
\usepackage{natbib}  
\usepackage{caption} 
\usepackage{multirow}
\usepackage{arydshln}
\usepackage{amsmath}
\usepackage{amsfonts}
\usepackage{booktabs}
\usepackage{algorithm}
\usepackage{algorithmic}
\usepackage{fontawesome}
\usepackage{newfloat}
\usepackage{listings}
\DeclareCaptionStyle{ruled}{labelfont=normalfont,labelsep=colon,strut=off} 
\floatstyle{ruled}
\newfloat{listing}{tb}{lst}{}
\floatname{listing}{Listing}
\title{Ethereum Fraud Detection via Joint Transaction Language Model and Graph Representation Learning}
\author{
    Jianguo Sun, Yifan Jia, Yanbin Wang, Yiwei Liu, Zhang Sheng, Ye Tian
}

\title{Ethereum Fraud Detection via Joint Transaction Language Model and Graph Representation Learning}
\author {
    Jianguo Sun, Yifan Jia, Yanbin Wang\textsuperscript{\faEnvelope}, Yiwei Liu, Zhang Sheng, Ye Tian
}

\begin{document}

\maketitle

\begin{abstract}
Ethereum faces growing fraud threats. Current fraud detection methods, whether employing graph neural networks or sequence models, fail to consider the semantic information and similarity patterns within transactions. Moreover, these approaches do not leverage the potential synergistic benefits of combining both types of models. To address these challenges, we propose TLMG4Eth that combines a transaction language model with graph-based methods to capture semantic, similarity, and structural features of transaction data in Ethereum. We first propose a transaction language model that converts numerical transaction data into meaningful transaction sentences, enabling the model to learn explicit transaction semantics. Then, we propose a transaction attribute similarity graph to learn transaction similarity information, enabling us to capture intuitive insights into transaction anomalies. Additionally, we construct an account interaction graph to capture the structural information of the account transaction network. We employ a deep multi-head attention network to fuse transaction semantic and similarity embeddings, and ultimately propose a joint training approach for the multi-head attention network and the account interaction graph to obtain the synergistic benefits of both.
\end{abstract}

%

\section{Introduction}

Blockchain technology has revolutionized various industries by providing a decentralized and secure method for recording transactions \cite{intro_r2}. Among blockchain platforms, Ethereum stands out for its robust support of smart contracts and decentralized applications \cite{intro_r3}. By introducing the concept of a programmable blockchain, Ethereum enabled developers to create applications beyond basic financial transactions \cite{intro_r1}.         

The growing popularity and value of Ethereum have attracted malicious actors, leading to an increase in phishing and fraud. According to The Chainalysis 2023 Crypto Crime Report\cite{chainalysis_report}, USD 39.6 billion worth of crypto-assets were received by identified illicit addresses, accounting for 0.42\% of total on-chain transaction volume—a significant increase from the previous year's USD 23.2 billion.

Ethereum fraud detection primarily relies on analyzing historical transaction records of accounts. Current methods predominantly employ Graph Neural Networks (GNNs) to model account transaction networks, or utilize sequence models such as Transformers \cite{attention} to process transaction histories. GNNs excel at capturing complex relationships and structural patterns within transaction networks, while sequence models are proficient in discerning temporal patterns and evolving behaviors.

However, current research fails to consider several crucial aspects: (1) Transaction Semantics: Existing approaches rely on historical transaction data in its numerical form, which lacks the context to interpret underlying intentions. As a result, the explicit meaning of transactions remains obscure, hindering models from understanding transaction semantics beyond mere numbers. (2) Transactional Similarity: Extracting similarity information from transaction attributes (such as amount, direction, and timing) is crucial for distinguishing between normal and anomalous transactions. Previous studies have overlooked the modeling of attribute similarities, which can offer direct insights into fraudulent behavior patterns. (3) Synergistic Optimization: While some studies have attempted to combine GNNs with sequence models, they typically adopt a late fusion approach, training the models separately and concatenating their features at the final stage. This approach fails to fully realize the potential synergies between the two methods.

To address the current challenges, we propose TLMG4Eth, which combines a transaction language model (TLM) with two transaction graphs to improve Ethereum fraud detection. We first train a transaction language model to learn explicit transaction semantics from transaction sentences, where transaction attributes (amount, direction, time, and other numerical data) are represented as words. Next, we propose a transaction attribute graph to model global semantic similarities between transactions and construct an account interaction graph to model transaction behaviors between accounts. We fuse transaction semantics, similarities, and structural information through a two-stage approach. First, we use a deep multi-head attention network to fuse the semantic embeddings and similarity embeddings of transactions. Then, we propose to jointly train the multi-head attention network with the account interaction graph to leverage their synergistic benefits.

Our main contributions include:
\begin{itemize}
\item We propose a transaction language model that transforms numerical transaction sequences into transaction sentences, clearly expressing transaction content and enabling the learning of explicit transaction semantics.
\item We propose a transaction attribute similarity graph to model global semantic similarities between transactions, thereby capturing intuitive insights into transaction anomalies.
\item We use a multi-head attention network to fuse transaction semantic and similarity information. Furthermore, we propose jointly training this multi-head attention network with an account interaction graph to obtain the benefits of both.
\item Our proposed method significantly outperforms current state-of-the-art approaches, improving F1-Scores by 10\%-20\% across three datasets.
\item We release a new dataset.
\end{itemize}

\section{RELATE WORK}

\subsection{Graph-based Methods} Graph-based methods construct transaction networks between accounts and employ graph embedding algorithms or GNNs for model training. \cite{node2vec,2vec_r1,2vec_r2} use Node2Vec to extract features from Ethereum transaction networks for fraudulent account detection. Trans2Vec \cite{2vec_r3} utilizes DeepWalk \cite{deepwalk}, dividing representation learning into node, edge, and attribute learning for classification. TGC \cite{GNN_r3} employs subgraph contrastive learning with statistical data for phishing address identification. PDGNN \cite{GNN_r4} proposes an end-to-end Chebyshev GCN model, using dynamic graph sampling and node information aggregation. MCGC \cite{GNN_r5} introduces a multi-channel graph classification model, automatically extracting information from different pooled graphs for phishing detection.

\subsection{Sequence-based Methods}
Graph-based approaches may struggle with high-frequency, repetitive transactions and long-term temporal patterns, leading some researchers to adopt sequence models that treat transactions as time-ordered event streams. BERT4ETH \cite{bert4eth,zipzap} exemplifies this approach, employing a BERT-like structure with a Transformer architecture to process chronological transaction events. It uses a masked language model for pre-training by randomly masking transaction addresses, then fine-tunes a Multi-Layer Perceptron (MLP) network for account classification.

\subsection{Hybrid Methods}
TSGN \cite{tsgn} proposes a transaction subgraph network model for phishing detection that integrates multiple feature representation methods, including Handcrafted and Diffpool techniques, while enhancing the classification model by introducing various mapping mechanisms within the transaction network. In a similar vein, TTAGN \cite{ttagn} employs a multi-step approach, initially utilizing Edge2Node to aggregate edge representations surrounding nodes in the transaction graph, followed by sampling time-based transaction aggregation graph networks from the transaction graph to learn temporal transaction sequences through LSTM; subsequently, it extracts statistical features from the raw data, ultimately concatenating these diverse features to perform classification.

\begin{figure*}[h]
    \centering
    \includegraphics[width=1\textwidth,height=0.51\textwidth]{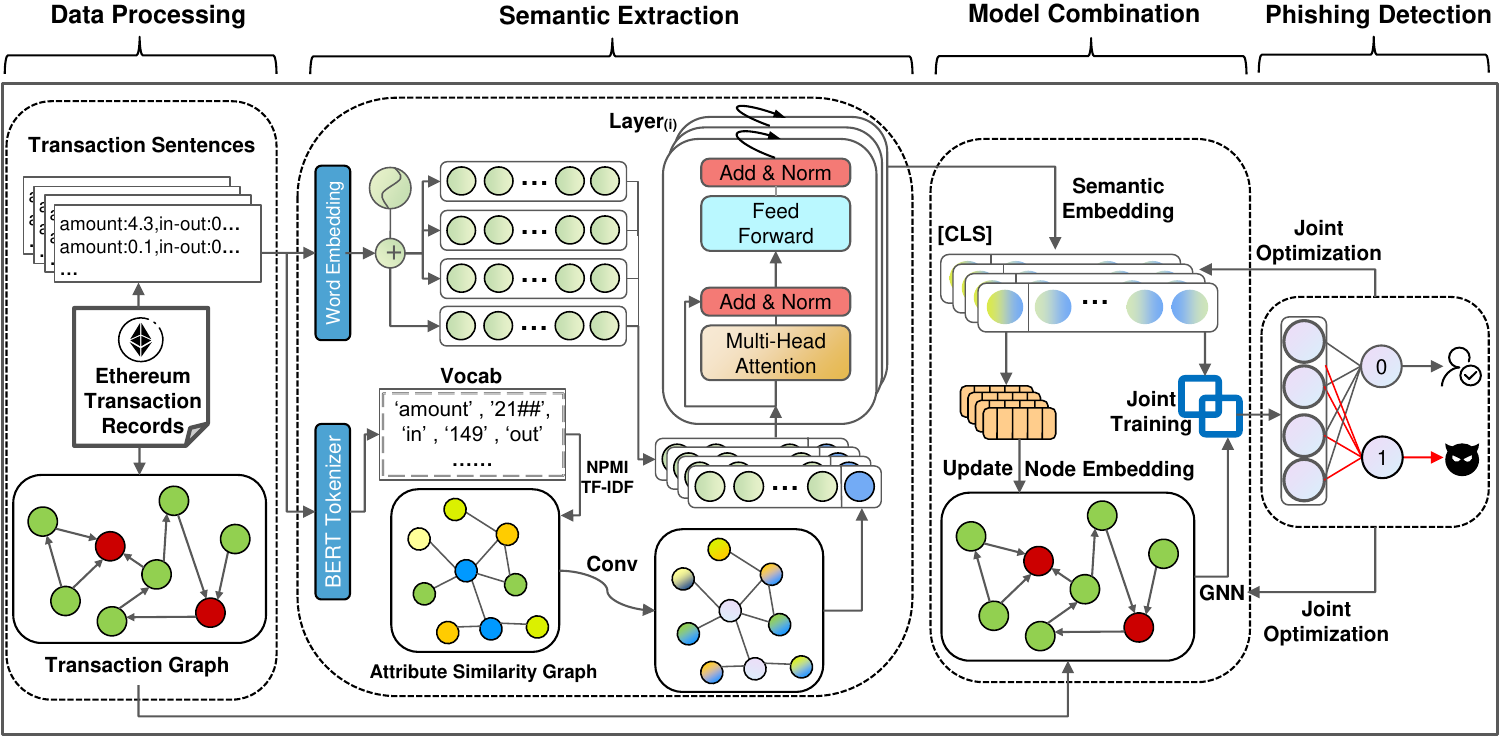}
    \caption{The framework of proposed Joint Transaction Language Model and Graph Representation Learning.}
    \label{fig:framwork}
\end{figure*}

\section{METHOD}
Figure \ref{fig:framwork} illustrates TLMG4Eth's architecture, whose key technical components are detailed below.

\subsection{Transaction Language Model}

The transaction language model(TLM) comprises two parts: first, we create a linguistic representation of numerical transaction data, then we employ a language model to extract semantic embeddings from transaction sentences.

\subsubsection{Linguistic Representation of Transactions} The numerical transaction data obscures specific transaction information. To address this, we propose a linguistic representation of transactions to elucidate their content. Let $\mathcal{T} = {t_1, t_2, ..., t_N}$ be a set of $N$ transactions for a single account. Each transaction $t_i$ is characterized by a tuple:

\begin{equation}
t_i = (v_i, d_i, \tau_i)
\end{equation}

where:
\begin{itemize}
    \item $d_i \in \{-1, 1\}$ is the direction, with -1 indicating inflow and 1 indicating outflow
    \item $\tau_i \in \mathbb{T}$ is the timestamp from the set of all possible timestamps $\mathbb{T}$
\end{itemize}
We transform each numerical attribute into a linguistic token by prepending a descriptive text indicator:

\begin{equation}
\mathcal{L}(t_i) = \{ \text{amount:} v_i, \text{direction:} d_i, \text{timetamp:} \tau_i\}
\end{equation}
Ethereum timestamps (e.g., 2024121214) lack interpretability and may mislead models due to their large numerical values. To address this, we propose to captures the intervals between N consecutive transactions rather than using raw timestamps. Let $\tau_i$ denote the timestamp of transaction $t_i$. The time intervals defined as:

\begin{equation}
\Delta\tau_{i,n} = \tau_i - \tau_{i-n}, \quad \text{for } n \in \{1, 2, ..., N-1\}
\end{equation}
Where $\Delta\tau_{i,n}$ represents the time difference between transaction $t_i$ and its $n$-th preceding transaction. We incorporate the time differences from the 2nd to the 5th preceding transactions into into $\mathcal{L}(t_i)$. The enhanced representation $\mathcal{L}'(t_i)$ is defined as:

\begin{equation}
\begin{split}
\mathcal{L}(t_i) = \{ & \text{amount:} v_i, \text{direction:} d_i, \\
                      & \text{2-inter\_time:} \Delta\tau_{i,2}, ..., \\
                      & \text{5-inter\_time:} \Delta\tau_{i,5}\}
\end{split}
\end{equation}

This enhanced representation captures transaction clustering at different time granularities. Each element in $\mathcal{L}(t_i)$ is treated as a transaction word. $N$ transactions of an account form a series of transaction sentences $\mathcal{C}$:

\begin{equation}
\mathcal{C} = \{\mathcal{L}(t_1), \mathcal{L}(t_2), ..., \mathcal{L}(t_N)\}
\end{equation}

\subsubsection{Transaction Semantic Embedding} We employ BERT-base \cite{bert} to extract semantic embeddings from these transaction representations. We continue training BERT using our domain-specific pre-training corpus, denoted as $\mathcal{D}$.

\begin{equation}
\mathcal{D} = \bigcup_{a \in \mathcal{A}} \mathcal{C}_a
\end{equation}
where $\mathcal{A}$ is the set of all accounts in Table 1, and $\mathcal{C}_a$ is the transaction sentences for account $a$. 

We use a masked language model (MLM) as objective:

\begin{equation}
\mathcal{L}_{\text{MLM}} = \mathbb{E}_{t \sim \mathcal{C}} \left[ -\sum_{i \in \mathcal{M}} \log P(t_i | \tilde{t}) \right]
\end{equation}
where $\mathcal{M}$ is the set of masked token indices, $\tilde{t}$ is the masked version of transaction sentence $t$, and $P(t_i | \tilde{t})$ is the probability of predicting the original token $t_i$ given the masked context. Post-training, for each token $x_i$ in a transaction sentence $t$, BERT-base generates a semantic embedding vector:
\vspace{-5pt}
\begin{equation}
\mathbf{e}_i = \text{BERT}(x_i | t) \in \mathbb{R}^d
\end{equation}
where $d$ is the dimensionality of the embedding space.

\subsection{Transaction Attribute Similarity Graph}

Transaction Attribute Similarity Graph (TASG) captures global transaction correlations across Ethereum data, providing intuitive insights into anomalies. For instance, phishing accounts typically conduct large transactions within short intervals, exhibiting similar semantic features in amount and temporal patterns.

\begin{figure}[t]
    \includegraphics[width=0.475\textwidth,height=0.26\textwidth]{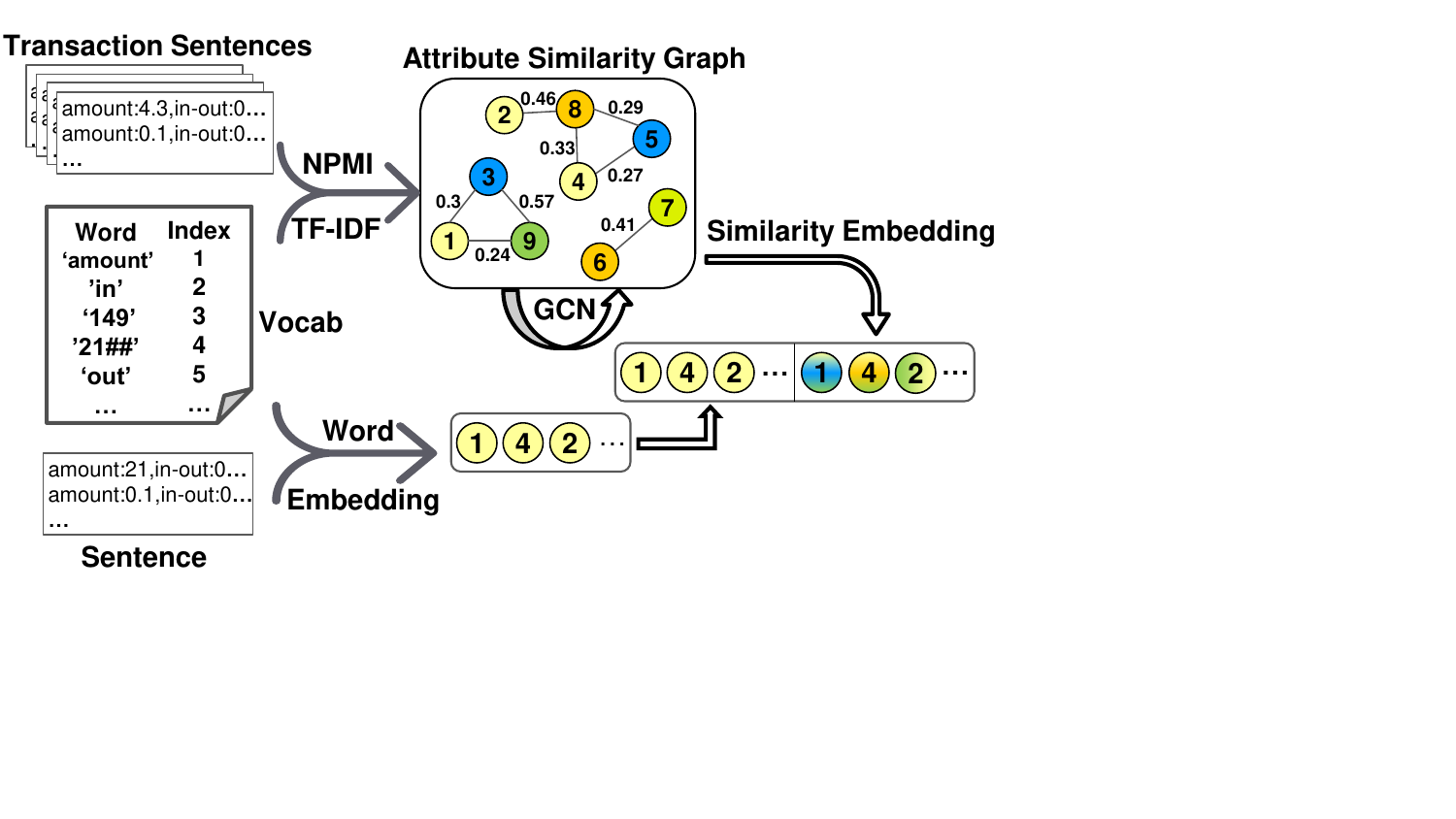}
    \caption{The generation and combination of ethereum transaction semantic embedding and similarity embedding.}
    \label{fig:framwork_1}
\end{figure}

We generate a vocabulary from the tokenized transaction corpus. Subsequently, we construct the TASG using two approaches: Normalized Pointwise Mutual Information (NPMI) and Term Frequency-Inverse Document Frequency (TF-IDF). Let TSSG be denoted as \(\mathcal{G}_w = (\mathcal{V}_w, \mathcal{E}_w)\), where \(\mathcal{V}_w\) represents the set of nodes corresponding to words in the vocabulary, and \(\mathcal{E}_w\) represents the set of edges connecting these nodes. The presence of an edge between two nodes is determined by either the NPMI or TF-IDF value.

The NPMI \cite{npmi} between two words $w_i$ and $w_j$ is calculated as follows:

\begin{equation}
\text{NPMI}(w_i, w_j) = \frac{\text{PMI}(w_i, w_j)}{-\log p(w_i, w_j)} = \frac{\log \frac{p(w_i, w_j)}{p(w_i) p(w_j)}}{-\log p(w_i, w_j)}
\end{equation}
where \( p(w_i, w_j) \) is the probability of co-occurrence of \( w_i \) and \( w_j \) within a given context window, \( p(w_i) \) and \( p(w_j) \) are the individual probabilities of \( w_i \) and \( w_j \). In our approach, the window size is the length of one transaction sentence, and we create an
edge between two words if their NPMI is larger than the predefined threshold \(\theta \).

The TF-IDF \cite{tfidf} between two words $w_i$ and $w_j$ is calculated as follows:

\vspace{-5pt}
\begin{equation}
\text{TF-IDF}(w_i, d) = \text{TF}(w_i, d) \times \log \left(\frac{N}{|\{d \in D : w_i \in d\}|}\right)
\end{equation}
where \( \text{TF}(w_i, d) \) is the term frequency of word \( w_i \) in transaction sentence \( d \), \( N \) is the total number of sentences in the corpus(all transaction records), \( |\{d \in D : w_i \in d\}| \) is the number of sentences containing \( w_i \). In our approach, we introduce additional sentence nodes in the vocabulary graph to model a TF-IDF based vocabulary graph \( \mathcal{G}_w\), the connectivity between sentence nodes and word nodes depends on whether their TF-IDF values exceed the predefined threshold \(\theta \).

We then apply graph convolution network(GCN) to encode nodes, obtaining global similarity embeddings for each word.

\begin{table}

    \renewcommand{\arraystretch}{1.1}
    \setlength{\tabcolsep}{3.7pt} 
    \centering
    \small
    \begin{tabular}{ccccc}  
        \toprule[1pt]
        \textbf{Dataset} & \textbf{Nodes} & \textbf{Edges} & \textbf{Avg Degree} & \textbf{Phisher}\\ \hline
        MulDiGraph & 2,973,489 & 13,551,303 & 4.5574 & 1,165\\ 
        B4E & 597,258 & 11,678,901 & 19.5542 & 3,220\\ 
        SPN & 496,740 & 831,082 & 1.6730 & 5,619 \\ 
        \bottomrule[1pt]
    \renewcommand{\arraystretch}{1}
    \end{tabular}
    \caption{Summary of three datasets.}
    \vspace{-15pt}
    \label{tab:dataset}
\end{table}

\subsection{Semantic and Similarity Embedding Fusion}

Since the vocabulary used for generating the transaction sequence semantic embeddings and constructing the vocabulary graph are both derived from the same tokenizer, the words in each account's transaction sequence are a subset of the vocabulary graph \cite{vgcn_bert}. As shown in Figure \ref{fig:framwork_1}, We select corresponding words from the TASG based on the input transaction sequence and concatenate the transaction similarity embeddings generated by TASG with the semantic embeddings generated by TLM. 

\begin{equation}
E_i = [E^s_i; E^g_i]
\end{equation}
Where $E^s_i$ is the semantic embedding for transaction $i$,$E^g_i$ is the similarity embedding from TASG for transaction $i$.

We fuse the information from the two types of embeddings in $E_i$ using a deep multi-head attention network (MAN) that consists of 12 layers, each with 12 attention heads. The computation for each attention head is as follows:
\vspace{-5pt}
\begin{equation}
\text{Attention}(\mathbf{Q}, \mathbf{K}, \mathbf{V}) = \text{softmax}\left(\frac{\mathbf{Q}\mathbf{K}^\top}{\sqrt{d_k}}\right)\mathbf{V}
\end{equation}

\begin{table*}[htbp]
\renewcommand{\arraystretch}{1.2}
\setlength{\tabcolsep}{3.7pt} 
\small 
\centering
\begin{tabular}{c|cccc|cccc|cccc}
\noalign{\hrule height 1pt} 
\multirow{2}[0]{*}{Method} & \multicolumn{4}{c|}{MulDiGraph} & \multicolumn{4}{c|}{B4E} & \multicolumn{4}{c}{SPN} \\ \cline{2-13}
&\textbf{Precision} & \textbf{Recall} & \textbf{F1} & \textbf{B-Acc} & \textbf{Precision} & \textbf{ Recall} & \textbf{F1} & \textbf{B-Acc} & \textbf{Precision} & \textbf{Recall} & \textbf{F1} & \textbf{B-Acc} \\ \hline
Role2Vec & 0.4688 & 0.6976 & 0.5608 & 0.6511 & 0.5748 & 0.7958 & 0.6673 & 0.7507 & 0.4521 & 0.7059 & 0.5512 & 0.6391 \\
Trans2Vec & 0.7114 & 0.6944 & 0.7029 & 0.7768 & 0.2634 & 0.7043 & 0.3842 & 0.3598 & 0.3928 & 0.7381 & 0.5134 & 0.5838 \\
GCN & 0.2960 & 0.7513 & 0.4247 & 0.4289 & 0.5515 & 0.7508 & 0.6359 & 0.7228 & 0.5046 & 0.4973 & 0.5009 & 0.6266 \\
GAT & 0.2689 & 0.7917 & 0.4014 & 0.3577 & 0.4729 & 0.8348 & 0.6038 & 0.6848 & 0.5083 & 0.7720 & 0.6130 & 0.6993 \\
SAGE & 0.3571 & 0.3299 & 0.3430 & 0.5164 & 0.4589 & 0.5826 & 0.5134 & 0.6196 & 0.4557 & 0.5817 & 0.5110 & 0.6172 \\
BERT4ETH & 0.4469 & 0.7344 & 0.5557 & 0.6400 & 0.7421 & 0.6125 & 0.6711 & 0.7530 & 0.7566 & 0.6713 & 0.7114 & 0.7817 \\ \hline
\textbf{Ours} & \textbf{0.8919 } & \textbf{0.9167 } & \textbf{0.9041 } & \textbf{0.9305 } & \textbf{ 0.8158 } & \textbf{ 0.8087 } & \textbf{ 0.8123 } & \textbf{ 0.8587 } & \textbf{ 0.7962 } & \textbf{ 0.8339 } & \textbf{ 0.8146 } & \textbf{ 0.8636 } \\ 
Improv. (\%) & 18.05 & 12.50 & 20.12 & 15.37 & 7.37 & -2.61 & 14.12 & 10.57 & 3.96 & 6.19 & 10.32 & 8.19 \\ 
\noalign{\hrule height 1pt} 
\end{tabular}%
\renewcommand{\arraystretch}{1}
\caption{The performances with our method and baseline methods on three datasets, and B-Acc is a Balanced Accuracy.}
\label{tab:baseline}%
\end{table*}%

\begin{figure*}[t]
    \centering
    \begin{minipage}[b]{0.095\textwidth}
        \centering
        \includegraphics[width=\textwidth]{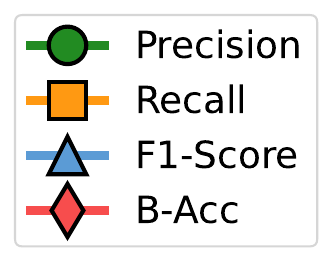}
        \vspace*{2.5mm}
    \end{minipage}
    \begin{minipage}[b]{0.19\textwidth}
        \centering
        \includegraphics[width=\textwidth]{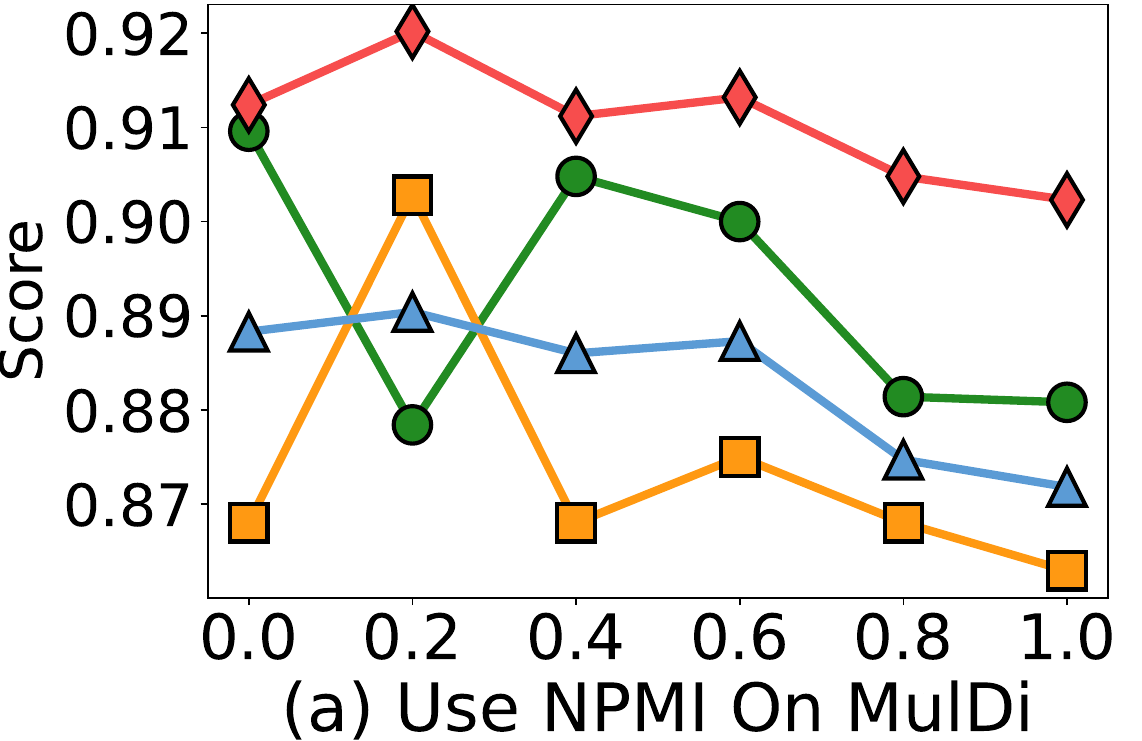}
    \end{minipage}
    \begin{minipage}[b]{0.19\textwidth}
        \centering
        \includegraphics[width=\textwidth]{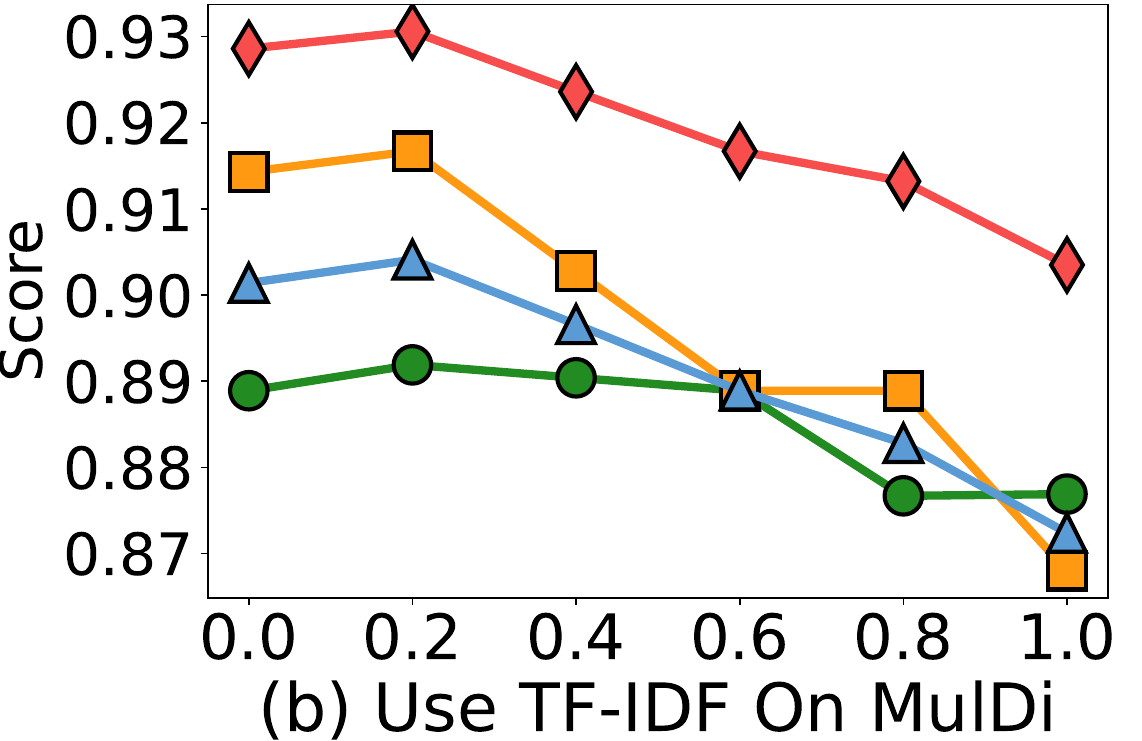}
    \end{minipage}
    \begin{minipage}[b]{0.19\textwidth}
        \centering
        \includegraphics[width=\textwidth]{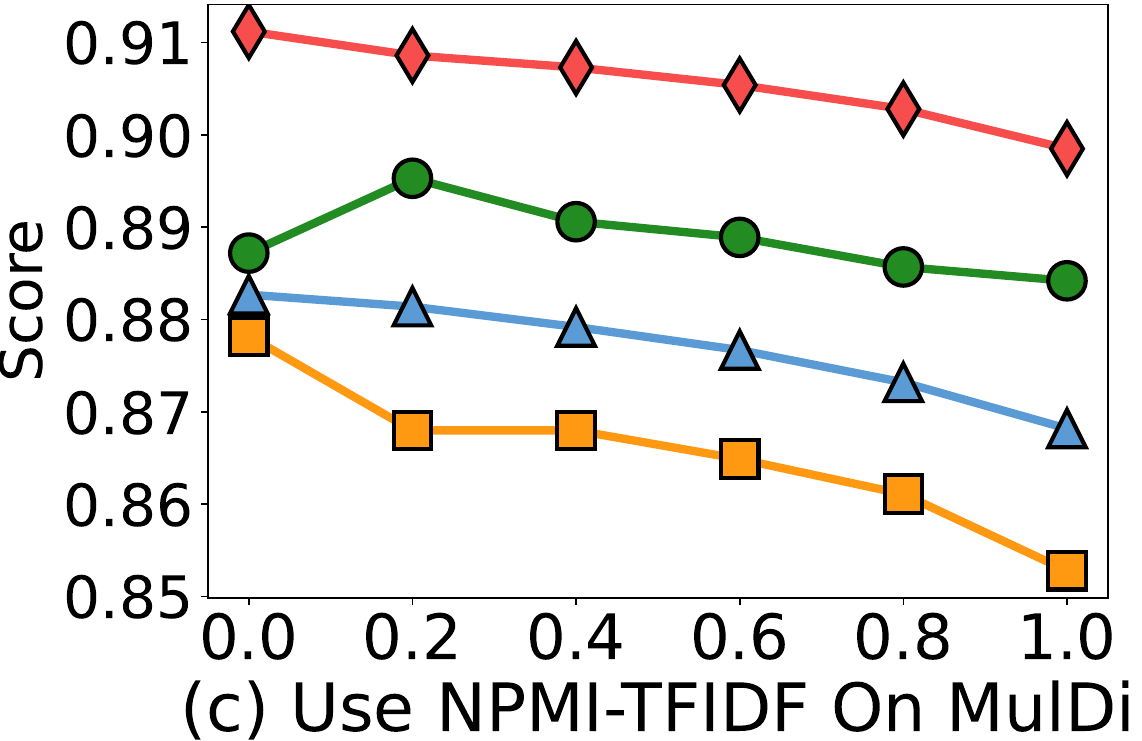}
    \end{minipage}
    \begin{minipage}[b]{0.19\textwidth}
        \centering
        \includegraphics[width=\textwidth]{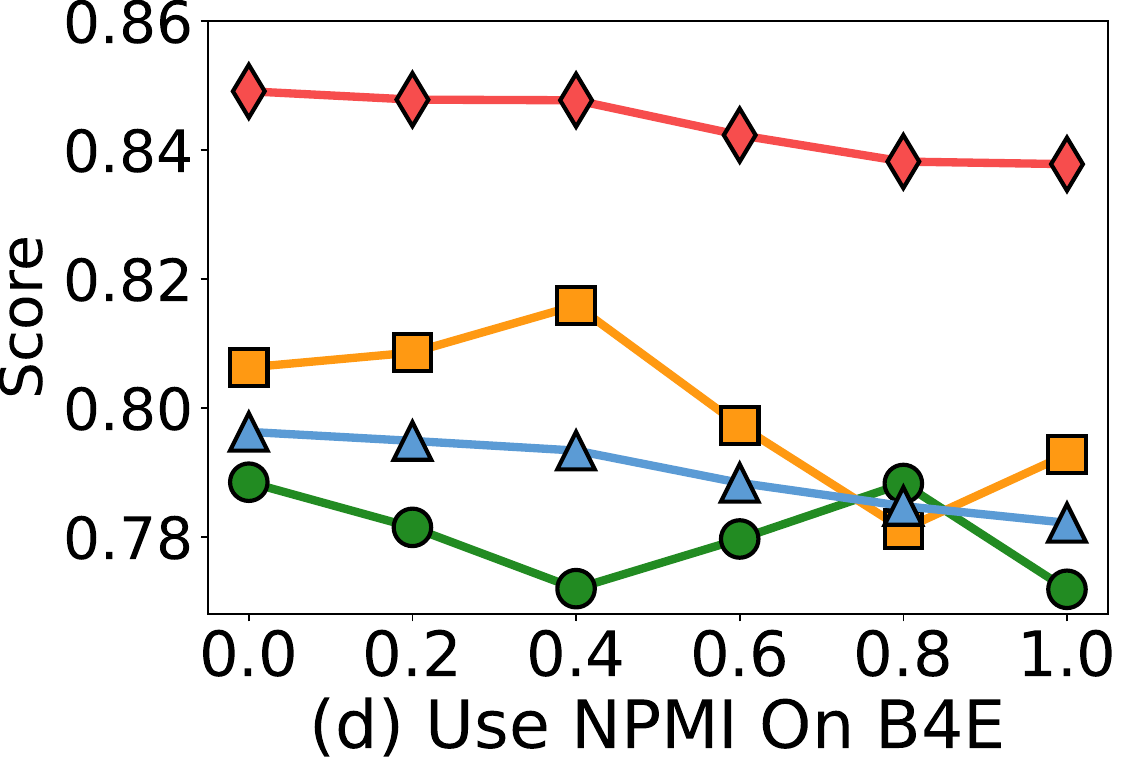}
    \end{minipage}
    \begin{minipage}[b]{0.095\textwidth}
        \centering
        \includegraphics[width=\textwidth]{crop_legend_only.pdf}
        \vspace*{2.5mm}
    \end{minipage}

    \begin{minipage}[b]{0.19\textwidth}
        \centering
        \includegraphics[width=\textwidth]{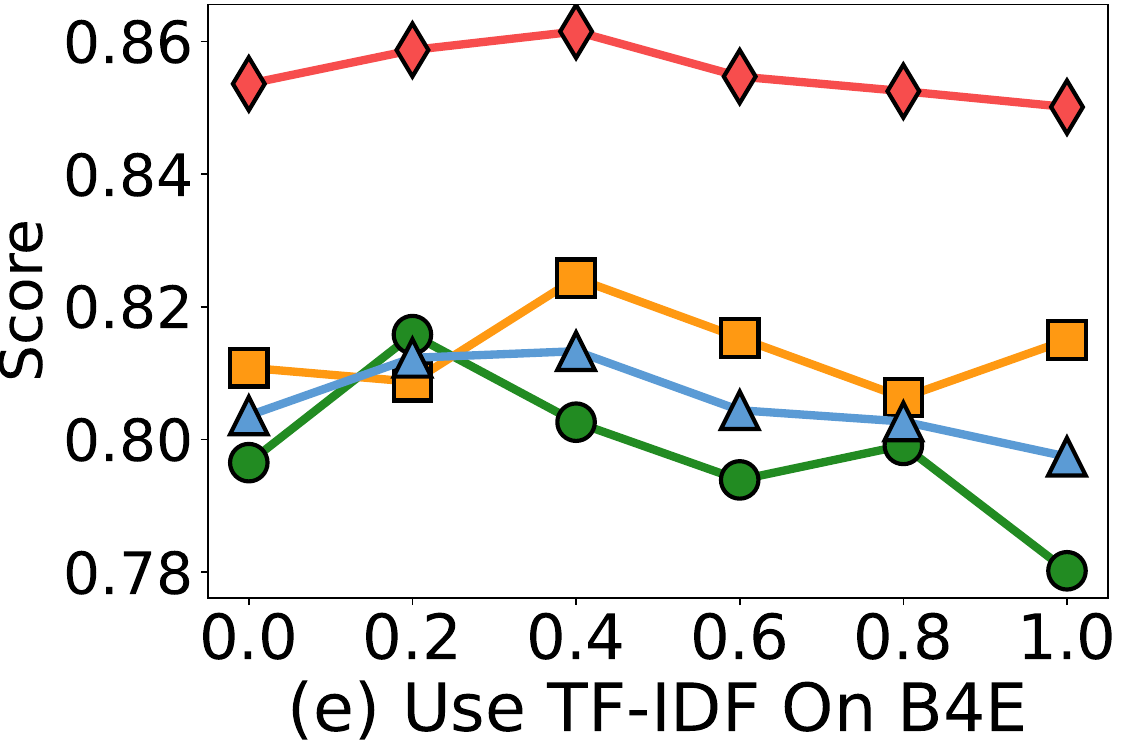}
    \end{minipage}
    \begin{minipage}[b]{0.19\textwidth}
        \centering
        \includegraphics[width=\textwidth]{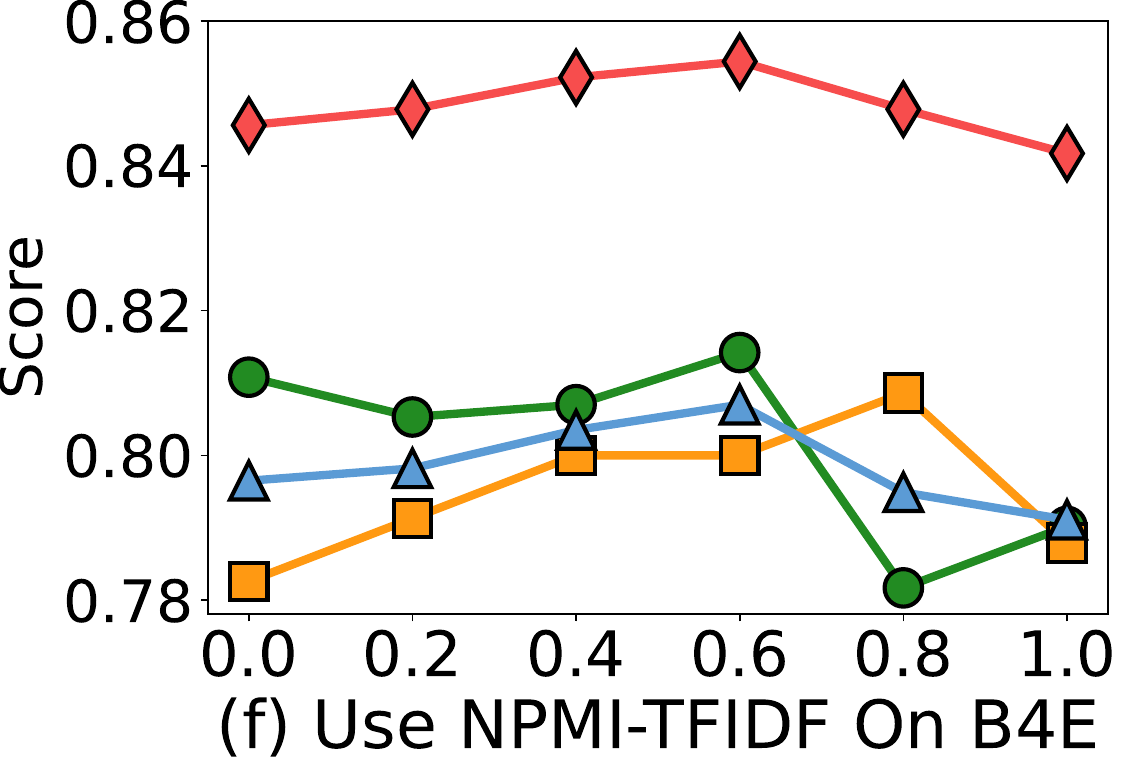}
    \end{minipage}
    \begin{minipage}[b]{0.19\textwidth}
        \centering
        \includegraphics[width=\textwidth]{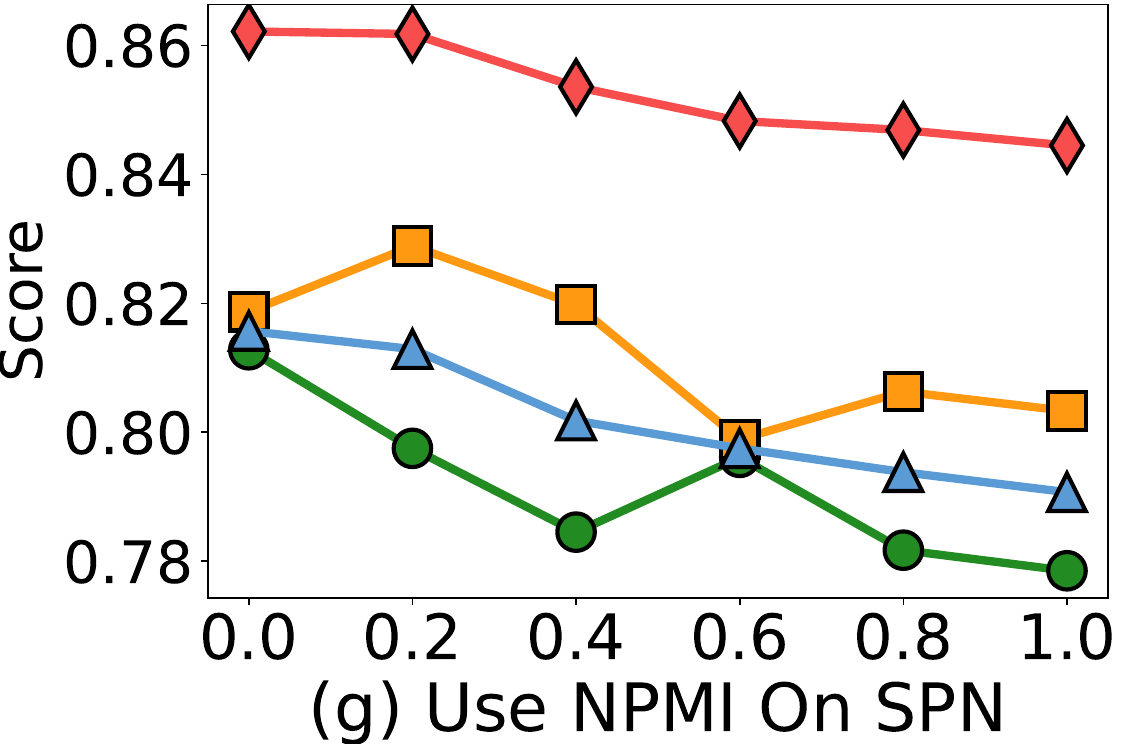}
    \end{minipage}
    \begin{minipage}[b]{0.19\textwidth}
        \centering
        \includegraphics[width=\textwidth]{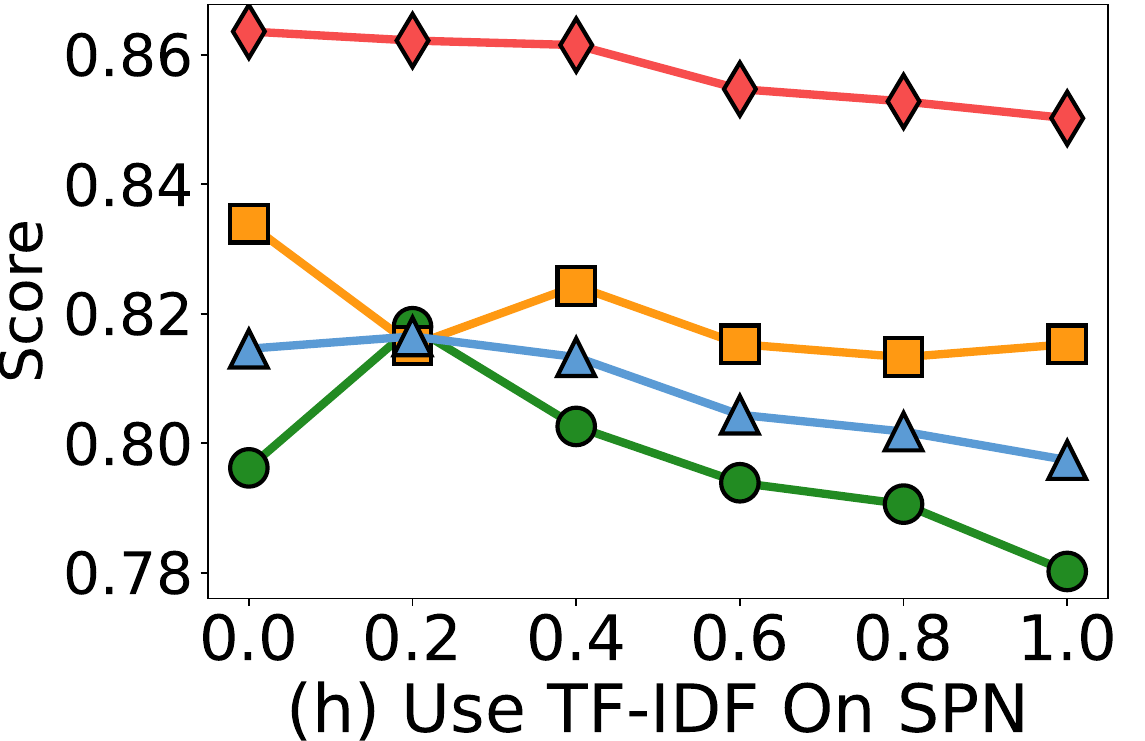}
    \end{minipage}
    \begin{minipage}[b]{0.19\textwidth}
        \centering
        \includegraphics[width=\textwidth]{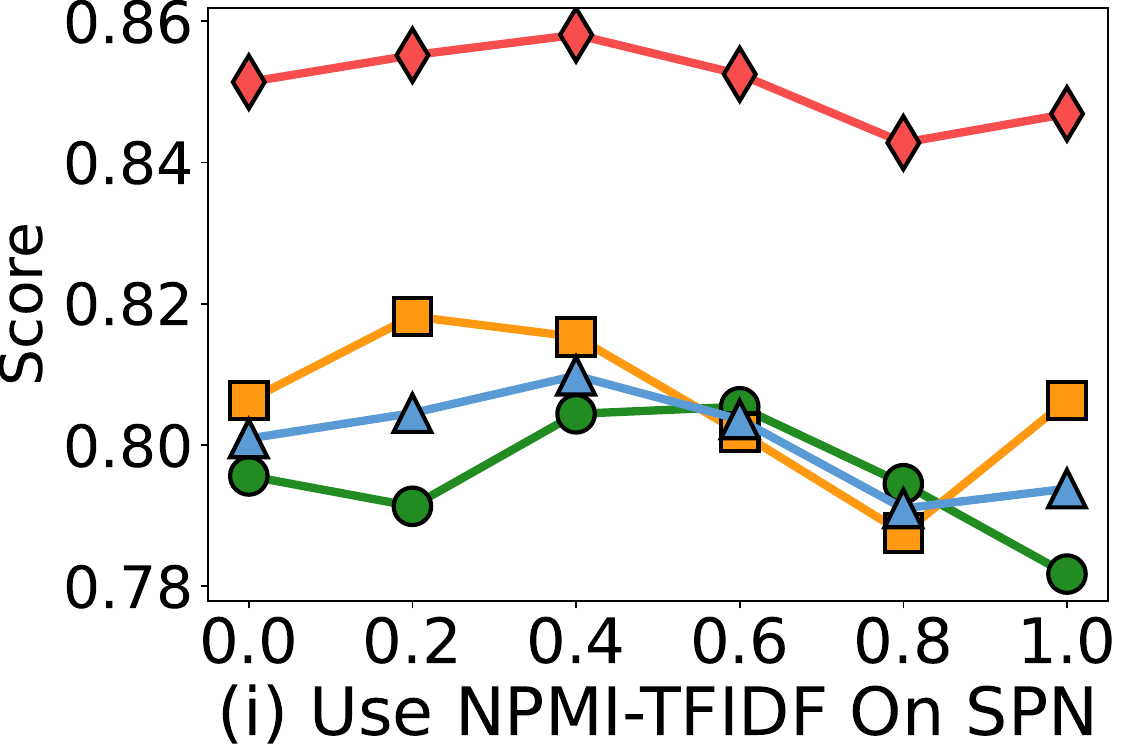}
    \end{minipage}

    \caption{Performance of various TASG construction methods under varying threshold $\theta$.}
    \label{fig:threshold}
\end{figure*}

\subsection{Account Interaction Graph}

We construct an Account Interaction Graph (AIG) to model transaction relationships between accounts and capture the topological information of transactions. We represent the AIG as a weighted graph \(G = (V, E)\),  where \( V \) is the set of account nodes in the transaction network, where each node represents an individual account. \( E \) is the set of edges, where each edge \( (i, j) \) represents that account \( i \) has transacted with account \( j \). The weight of an edge \( w_{ij} \) represents the number of transactions between account \( i \) and account \( j \).

The account nodes are iteratively updated using GCN, and update rule for layer \( l+1 \) is given by:

\begin{equation}
\mathbf{H}^{(l+1)} = \sigma\left( \mathbf{\hat{A}} \mathbf{H}^{(l)} \mathbf{W}^{(l)} \right)
\end{equation}
where \( \mathbf{H}^{(l)} \in \mathbb{R}^{N \times F^{(l)}} \) is the node feature matrix at layer \( l \), \( \mathbf{\hat{A}} = \mathbf{D}^{-\frac{1}{2}} (\mathbf{A} + \mathbf{I}) \mathbf{D}^{-\frac{1}{2}} \) is the normalized adjacency matrix with self-loops added (\( \mathbf{I} \) is the identity matrix and \( \mathbf{D} \) is the degree matrix), \( \mathbf{W}^{(l)} \in \mathbb{R}^{F^{(l)} \times F^{(l+1)}} \) is the trainable weight matrix for layer \( l \), \( \sigma(\cdot) \) is ReLU, the non-linear activation function.

\subsection{Joint Training of MAN and AIG}

We combine two approaches for joint model training. First, we use a multi-head attention network to initialize account interaction graph node embeddings, enabling collaborative training as the softmax function updates \cite{bertgcn,liu2024source}.
\begin{equation}
\mathbf{Z}_{\text{GCN}} = softmax(g(\mathbf{X}, \mathbf{A}))
\end{equation}

Where \(\mathbf{X}\) is the initial node embedding from the multi-head attention network.

Then, we linearly interpolate predictions from the MAN and GCN to obtain the final prediction:

\begin{equation}
\text{Pred} = \lambda \text{Z}_{\text{GCN}} + (1 - \lambda) \times \text{Z}_{\text{MAN}} 
\end{equation}

$\text{Z}_{\text{MAN}}$ represents the prediction of the multi-head attention network. The parameter $\lambda$ controls how much weight the two models contribute to the result. $\lambda$ = 0 means that we use the full GCN model, while $\lambda$ = 1 means that we just use the transaction language model.

\begin{table*}[t]
\renewcommand{\arraystretch}{1.2}
\setlength{\tabcolsep}{3.8pt} 
\small 
\centering
\begin{tabular}{c|cccc|cccc|cccc}
\noalign{\hrule height 1pt}
\multirow{2}[0]{*}{Enhancer} & \multicolumn{4}{c|}{MulDiGraph} & \multicolumn{4}{c|}{B4E} & \multicolumn{4}{c}{SPN} \\ \cline{2-13}
&\textbf{Precision} & \textbf{Recall} & \textbf{F1} & \textbf{B-Acc} & \textbf{Precision} & \textbf{ Recall} & \textbf{F1} & \textbf{B-Acc} & \textbf{Precision} & \textbf{Recall} & \textbf{F1} & \textbf{B-Acc} \\ \hline
    w/o   & 0.8776  & 0.8731  & 0.8753  & 0.9061  & 0.7807  & 0.7850  & 0.7825  & 0.8374  & 0.7902  & 0.7879  & 0.7911  & 0.8417  \\ \cdashline{1-13}
    NPMI-TFIDF & \textbf{0.8953 } & 0.8680  & 0.8814  & 0.9086  & 0.8053  & 0.7913  & 0.7982  & 0.8478  & 0.7913  & 0.8182  & 0.8045  & 0.8551  \\ 
    Improv. (\%) & \textbf{1.77} & -0.51 & 0.61 & 0.25 & 2.46 & 0.63 & 1.57 & 1.04 & 0.11 & 3.03 & 1.34 & 1.34 \\ \cdashline{1-13}
    NPMI  & 0.8784  & 0.9028  & 0.8904  & 0.9202  & 0.7815  & 0.8086  & 0.7949  & 0.8478  & \textbf{0.7975 } & 0.8289  & 0.8129  & 0.8618  \\
    Improv. (\%) & 0.08 & 2.97 & 1.51 & 1.40 & 0.08 & 2.36 & 1.24 & 1.04 & \textbf{0.73} & 4.10 & 2.18 & 2.01 \\ \cdashline{1-13}
    TF-IDF & 0.8919  & \textbf{0.9167 } & \textbf{0.9041 } & \textbf{0.9306 } & \textbf{0.8158 } & \textbf{0.8087 } & \textbf{0.8123 } & \textbf{0.8587 } & 0.7962  & \textbf{0.8339 } & \textbf{0.8146 } & \textbf{0.8636 } \\
    Improv. (\%) & 1.43 & \textbf{4.36} & \textbf{2.88} & \textbf{2.45} & \textbf{3.51} & \textbf{2.37} & \textbf{2.98} & \textbf{2.13} & 0.60 & \textbf{4.60} & \textbf{2.35} & \textbf{2.19} \\ 
    \noalign{\hrule height 1pt}
    \end{tabular}%
\renewcommand{\arraystretch}{1}
\caption{Performance comparison of TLM combined with different versions of TASG versus TLM alone.}
\label{tab:similarity}%
\end{table*}%

To reduce computational complexity and memory requirements, we introduce a batch update method to achieve synchronous mini-batch training for both the attention network and GCN. Specifically, we construct a dictionary to track the embeddings of all accounts in both models. In each iteration, we sample a mini-batch from phishing and normal accounts by first computing their semantic embeddings, then updating the corresponding node embeddings in the transaction graph network through the dictionary. Next, we use the updated semantic embeddings of the nodes to derive the GCN output, calculate the cross-entropy loss for the current mini-batch after performing prediction interpolation. The loss function can be expressed as follows.

\vspace{-10pt}
\begin{equation}
\mathcal{L} = - \sum_{i=1}^{2} y_i \log(\text{Pred}_i) + (1 - y_i) \log(1 - \text{Pred}_i)
\end{equation}

\section{DATASET REVIEW }
As shown in Table \ref{tab:dataset}, we utilized three datasets: MulDiGraph, B4E, and SPN.       

\subsubsection{MulDiGraph} This dataset is publicly available on the XBlock  platform \cite{xblock}.        It includes a large Ethereum transaction network obtained by performing a two-hop Breadth-First Search (BFS) from known phishing nodes.        The dataset contains 2,973,489 nodes, 13,551,303 edges, and 1,165 phishing nodes.

\subsubsection{B4E} This dataset was collected via an Ethereum node using Geth \cite{bert4eth}. It covers transactions from January 1, 2017, to May 1, 2022, including 3,220 phishing accounts and 594,038 normal accounts. The dataset contains 328,261 transactions involving phishing accounts and 1,350,640 involving normal accounts. It's divided into four groups: phishing accounts, normal accounts, incoming transactions, and outgoing transactions.

\subsubsection{Our Dataset SPN} Second order Phishing Network, we created this dataset using the Etherscan API. Starting from the most recently identified all phishing nodes, we performed a two-hop BFS to gather information on their neighbors and extracted the first 100 transaction records for each involved node \cite{xblock_1}. This dataset contains trading information prior to June 7, 2024, includes 5,619 phishing accounts and 491,121 normal accounts, with a total of 831,082 transaction edges. SPN provides the most up-to-date view of the Ethereum network's transaction landscape, reflecting recent phishing activities and network dynamics.

\section{EXPERIENCE}
TLMG4Eth is compared against several baselines including graph embedding methods (Role2Vec \cite{role2vec}, Trans2Vec \cite{2vec_r3}), graph neural networks (GCN \cite{GNN_r1}, GAT \cite{gat}, SAGEConv \cite{sage}), and a Transformer-based method (BERT4ETH \cite{bert4eth}). For practical considerations, we limit our analysis to the 100 most recent transactions per account. Our model configuration employs the BERT-base architecture for the language model, with both the transaction attribute similarity graph and account interaction graph utilizing 2 GCN layers, while joint training uses a batch sampling size of 64 and a learning rate of 1e-5. For baseline methods, we maintain the original parameter settings for Trans2Vec and BERT4ETH as reported in their respective publications, while other baselines retain their standard configurations without modifications.

\subsection{Comparison with Baselines}

Table \ref{tab:similarity} presents the comparative results of TLMG4Eth and six baseline methods. TLMG4Eth notably outperforms all baseline models across the three datasets. Specifically, TLMG4Eth surpasses the strongest baseline in F1-Scores by approximately 20.12\%, 14.12\%, and 10.32\% on MudiGraph, B4E, and SPN respectively. Similarly, for Balanced Accuracy, it shows improvements of 15.37\%, 10.57\%, and 8.19\% respectively.
Notably, TLMG4Eth's performance gains on MudiGraph and SPN datasets are significantly higher compared to those on the B4E dataset. This disparity may be attributed to the data collection methodology employed for MudiGraph and SPN. These datasets incorporated network structure considerations when extracting information from the Ethereum blockchain, utilizing breadth-first search algorithms initiated from phishing nodes. This approach likely facilitated the formation of larger connected graphs during the construction of Ethereum transaction networks, thereby enabling the model to capture more comprehensive global and topological information.

Furthermore, graph embedding-based algorithms demonstrate high recall rates across all datasets but exhibit suboptimal precision. This suggests an overemphasis on positive samples (phishing nodes) at the expense of learning from negative samples, resulting in elevated false positive rates. Conversely, graph neural network-based models aggregate more features from neighboring normal accounts, leading to increased attention on regular nodes (negative samples). Consequently, GNN models tend to experience diminished classification performance for phishing nodes during the training process.
\begin{table*}[t]
  \centering
  \renewcommand{\arraystretch}{1.2}
  \setlength{\tabcolsep}{4.6pt} 
  \small 
    \begin{tabular}{c|cccc|cccc|cccc}
    \noalign{\hrule height 1pt}
    \multirow{2}[0]{*}{$\lambda$} & \multicolumn{4}{c|}{MulDiGraph} & \multicolumn{4}{c|}{B4E}       & \multicolumn{4}{c}{SPN} \\ \cline{2-13}
          & \textbf{Precision} & \textbf{Recall } & \textbf{F1} & \textbf{B-Acc} & \textbf{Precision} & \textbf{Recall } & \textbf{F1} & \textbf{B-Acc} & \textbf{Precision} & \textbf{Recall } & \textbf{F1} & \textbf{B-Acc} \\ \hline
    0     & 0.8918  & 0.8782  & 0.8849  & 0.9125  & 0.7648  & 0.8003  & 0.7820  & 0.8386  & 0.7856  & 0.8103  & 0.7972  & 0.8499  \\ \cdashline{1-13}
    0.1   & 0.8788  & 0.8832  & 0.8810  & 0.9111  & 0.7741  & 0.8072  & 0.7903  & 0.8447  & 0.7949  & 0.8142  & 0.8040  & 0.8546  \\
    0.2   & 0.8964  & 0.8782  & 0.8872  & 0.9137  & \underline{\textbf{ 0.7963 }} & 0.7822  & 0.7892  & 0.8411  & \underline{\textbf{ 0.8174 }} & 0.7992  & 0.8082  & 0.8550  \\
    0.3   & \underline{\textbf{ 0.9014 }} & 0.8889  & 0.8951  & 0.9201  & 0.7716  & 0.8163  & 0.7933  & 0.8477  & 0.7923  & 0.8333  & 0.8123  & 0.8620  \\
    0.4   & 0.9000  & 0.8750  & 0.8873  & 0.9132  & 0.7702  & 0.8134  & 0.7912  & 0.8460  & 0.7910  & 0.8304  & 0.8102  & 0.8603  \\
    0.5   & 0.8974  & 0.8883  & 0.8929  & 0.9188  & 0.7875  & 0.7885  & 0.7878  & 0.8411  & 0.8127  & 0.8055  & 0.8091  & 0.8563  \\
    0.6   & 0.9028  & 0.9028  & 0.9028  & 0.9271  & 0.7445  & \underline{\textbf{0.8259 }} & 0.7830  & 0.8421  & 0.7693  & 0.8429  & 0.8044  & 0.8582  \\
    0.7   & 0.8919  & \underline{\textbf{ 0.9167 }} & \underline{\textbf{ 0.9041 }} & \underline{\textbf{ 0.9306 }} & 0.7824  & 0.7947  & 0.7885  & 0.8421  & 0.8033  & 0.8117  & 0.8075  & 0.8562  \\
    0.8   & 0.8904  & 0.9028  & 0.8966  & 0.9236  & 0.7817  & 0.8086  & \underline{\textbf{0.7949 }} & \underline{\textbf{ 0.8478 }} & 0.7821  & \underline{\textbf{ 0.8491 }} & \underline{\textbf{ 0.8142 }} & \underline{\textbf{ 0.8654 }} \\
    0.9   & 0.8767  & 0.8889  & 0.8828  & 0.9132  & 0.7676  & 0.8026  & 0.7847  & 0.8406  & 0.7712  & 0.8397  & 0.8040  & 0.8576  \\ \cdashline{1-13}
    1     & 0.3172  & 0.8194  & 0.4574  & 0.4687  & 0.5742  & 0.7477  & 0.6495  & 0.7352  & 0.5742  & 0.7477  & 0.6495  & 0.7352  \\ 
    \noalign{\hrule height 1pt}
    \end{tabular}%
    \renewcommand{\arraystretch}{1}
    \caption{Effect of varying trade-off parameter $\lambda$ between TLM and AIG on model. }
  \label{tab:tradeoff}%
\end{table*}%

\begin{figure}[t]
    \centering
    \begin{minipage}[b]{0.152\textwidth}
        \centering
        \includegraphics[width=\textwidth]{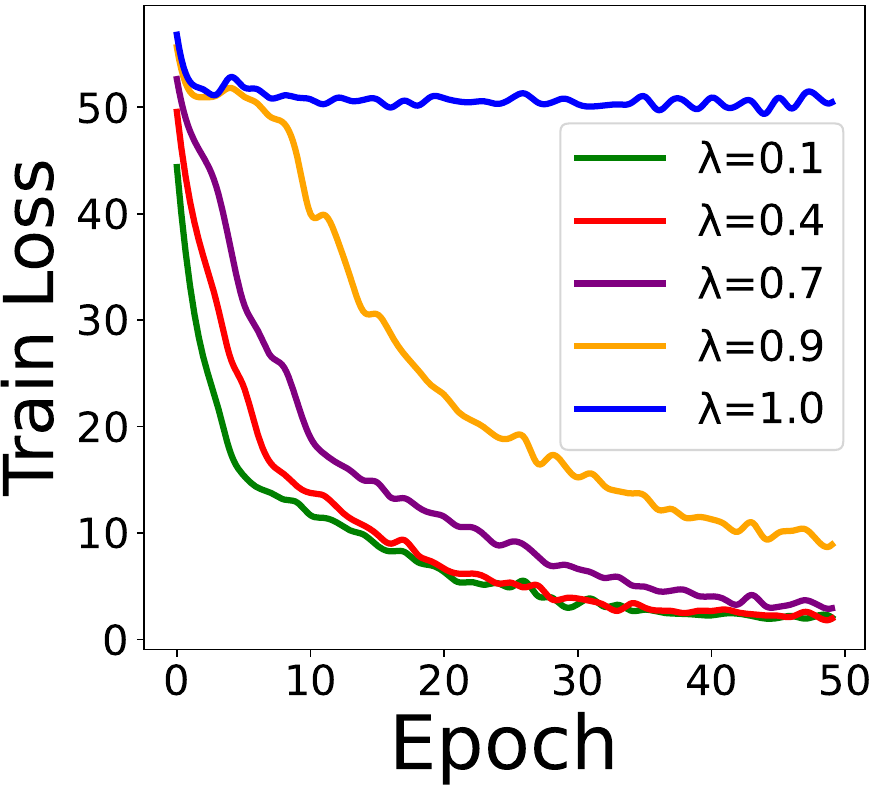}
    \end{minipage}
    \begin{minipage}[b]{0.152\textwidth}
        \centering
        \includegraphics[width=\textwidth]{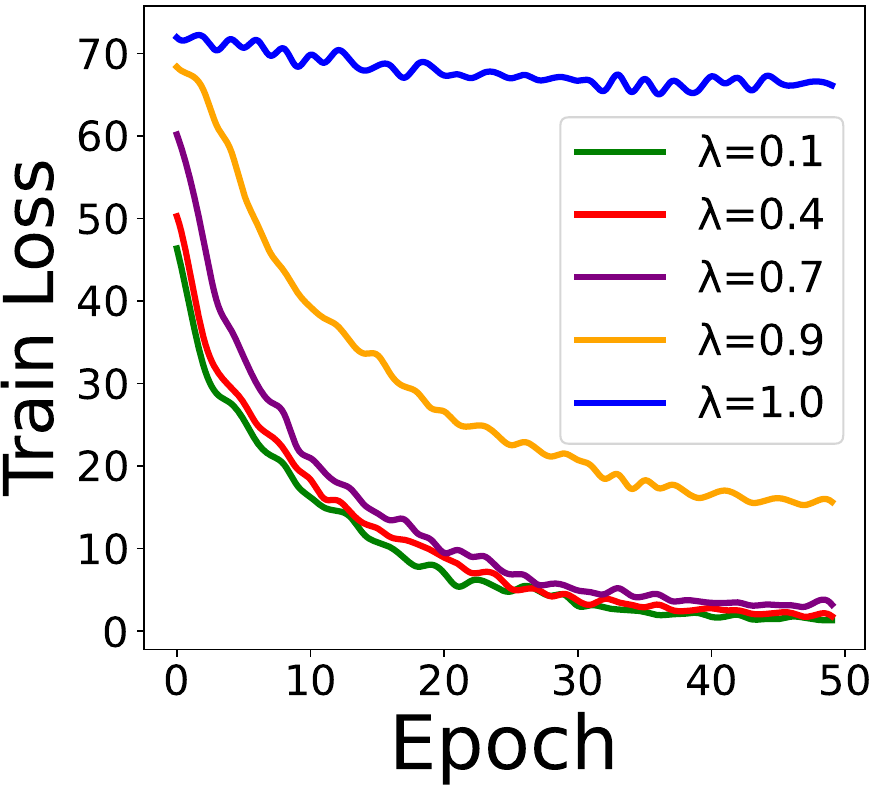}
    \end{minipage}
    \begin{minipage}[b]{0.152\textwidth}
        \centering
        \includegraphics[width=\textwidth]{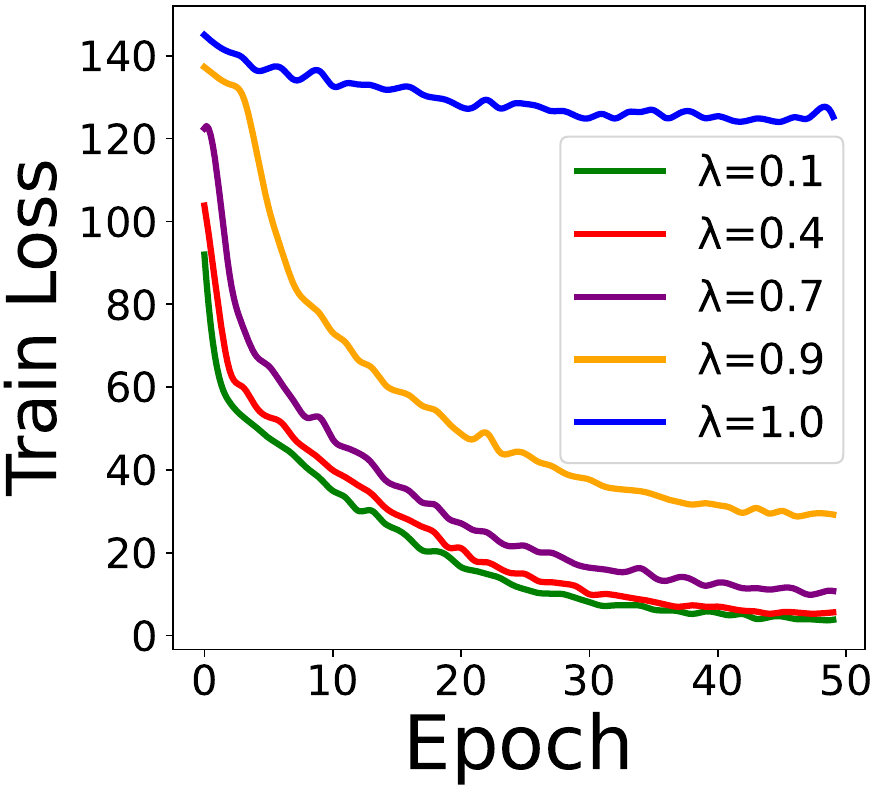}
    \end{minipage}
    \caption{Training Loss vs Epoch on MulDiGraph, B4E and SPN datasets with different trade-off parameter $\lambda$.}
    \label{fig:train_loss}
    \vspace{-10pt}
\end{figure}

\subsection{Ablation Study of Attribute Similarity Graph}

\begin{table*}[h]
  \renewcommand{\arraystretch}{1.2}
  \setlength{\tabcolsep}{3.8pt} 
  \small 
  \centering
    \begin{tabular}{c|cccc|cccc|cccc}
    \noalign{\hrule height 1pt}
    \multirow{2}[0]{*}{Combination} & \multicolumn{4}{c|}{MulDiGraph} & \multicolumn{4}{c|}{B4E}       & \multicolumn{4}{c}{SPN} \\ \cline{2-13}
          & \textbf{Precision} & \textbf{Recall } & \textbf{F1} & \textbf{B-Acc} & \textbf{Precision} & \textbf{Recall } & \textbf{F1} & \textbf{B-Acc} & \textbf{Precision} & \textbf{Recall } & \textbf{F1} & \textbf{B-Acc} \\ \hline
    TLM Only   & 0.8827  & 0.8782  & 0.8804  & 0.9099  & 0.8002  & 0.7841  & 0.7923  & 0.8431  & 0.7980  & 0.8056  & 0.8018  & 0.8518  \\ \cdashline{1-13}
    TLM+GCN & 0.8667  & 0.9028  & 0.8844  & 0.9167  & 0.8074  & 0.7974  & 0.8027  & 0.8511  & 0.7905  & 0.8128  & 0.8015  & 0.8526  \\
    Improv. (\%) & -1.60  & 2.46  & 0.40  & 0.68  & 0.72  & 1.33  & 1.04  & 0.80  & -0.75  & 0.72  & -0.03  & 0.07  \\ \cdashline{1-13}
    TLM+GAT & 0.8767  & 0.8889  & 0.8828  & 0.9132  & 0.8086  & 0.7818  & 0.7949  & 0.8446  & \textbf{0.8107 } & 0.7984  & 0.8045  & 0.8526  \\
    Improv. (\%) & -0.60  & 1.07  & 0.24  & 0.33  & 0.84  & -0.23  & 0.26  & 0.15  & \textbf{1.27 } & -0.72  & 0.27  & 0.08  \\ \cdashline{1-13}
    TLM+SAGE & 0.8866  & 0.8731  & 0.8798  & 0.9086  & 0.8032  & 0.8037  & 0.8032  & 0.8526  & 0.7813  & 0.8182  & 0.7993  & 0.8518  \\
    Improv. (\%) & 0.39  & -0.51  & -0.06  & -0.13  & 0.30  & 1.96  & 1.09  & 0.95  & -1.68  & 1.26  & -0.25  & 0.00  \\ \cdashline{1-13}
    TLM+GCNe & \textbf{0.8919 } & \textbf{0.9167 } & \textbf{0.9041 } & \textbf{0.9306 } & \textbf{0.8158 } & \textbf{0.8087 } & \textbf{0.8123 } & \textbf{0.8587 } & 0.7962  & \textbf{0.8339 } & \textbf{0.8146 } & \textbf{0.8636 } \\
    Improv. (\%) & \textbf{0.92 } & \textbf{3.85 } & \textbf{2.37 } & \textbf{2.06 } & \textbf{1.56 } & \textbf{2.46 } & \textbf{2.00 }  & \textbf{1.56 } & -0.19  & \textbf{2.83 } & \textbf{1.28 } & \textbf{1.17 } \\ 
    \noalign{\hrule height 1pt}
    \end{tabular}%
    \caption{Performance of TLM combined with various GNN.}
  \label{tab:diff_gnn}%
\end{table*}%

In this section, we examine the impact of transaction attribute similarity graphs on model prediction performance. NPMI refers to the use of Normalized Pointwise Mutual Information to construct transaction attribute similarity graphs, while TF-IDF indicates the use of Term Frequency-Inverse Document Frequency. NPMI-TFIDF represents a combination of NPMI and TF-IDF approaches. The notation "w/o" signifies the absence of transaction attribute similarity information.
The experimental results are presented in Figure \ref{fig:threshold} and Table \ref{tab:similarity}. Overall, regardless of the communication method employed, the introduction of transaction attribute similarity graphs improved model performance. Notably, the TF-IDF-based attribute similarity graph yielded significant improvements, increasing the F1-Scores by approximately 2.88\%, 2.98\%, and 2.35\% for MulDiGraph, B4E, and SPN, respectively. The B-Acc metric also saw improvements of about 2.45\%, 2.13\%, and 2.19\% across these models.
This enhanced performance can be attributed to the nature of Ethereum transaction scenarios, where phishing accounts and their transaction records appear frequently in phishing scams but constitute a small proportion of total transaction information. Consequently, words associated with phishing in transaction language have higher TF-IDF values, effectively capturing key information about phishing accounts.
NPMI-based methods focus on co-occurrence probabilities. PMI can be understood as a pre-clustering of corpus information, where strongly correlated words are grouped together. In the context of normal Ethereum account transactions, most transaction records are similar, leading PMI to focus more on the semantics of normal account transactions. Therefore, compared to TF-IDF information, PMI provides limited similarity features for phishing accounts.
Furthermore, the improvements brought by attribute similarity graphs are more pronounced when the threshold $\theta$ is relatively low, between 0 and 0.4. This is because higher $\theta$ values filter out edges for most words, reducing the generalizability of similarity information.

\subsection{Impact of the Trade-off Parameter}

In this section, we discuss the impact of the trade-off parameter $\lambda$ on the performance of the jointly trained model. The experimental results are presented in Table \ref{tab:tradeoff}.
Across all three datasets, the results of joint training outperform the cases where $\lambda$ = 0 or 1. This indicates that joint training is more effective than using either approach independently. Specifically, the model achieves peak F1-Scores and B-Acc when $\lambda$ is set to 0.7, 0.8, and 0.8 for the respective datasets. The highest F1-Scores attained are 90.41\%, 79.49\%, and 81.42\%, while the highest B-Acc values are 93.06\%, 84.78\%, and 86.54\%. We also compared the training loss curves of the model under different $\lambda$ values. As illustrated in Figure \ref{fig:train_loss}, when $\lambda$ is set to 0.9, the model's training loss converges to a higher value compared to other $\lambda$ settings. When $\lambda$ is set to 1, the loss barely decreases as training progresses, resulting in a significant drop in model performance, to the point where it is comparable to the GCN performance in the baseline methods.

\subsection{Different GNN Model Combination}

In this section, we explore the performance differences of various graph representation learning algorithms applied to AIG. We evaluate four graph learning algorithms, each jointly trained with PLM, including: Graph Convolutional Network (GCN) without edge features, GCN with edge features (GCNe), Graph Attention Network (GAT), and GraphSAGE convolution (SAGEconv). For GCNe, we utilize the number of transactions between two nodes as the edge feature. All GNN models are configured with two layers, while GAT is set up with 8 attention heads, and SAGEconv employs mean aggregation.

The experimental results, presented in Table \ref{tab:diff_gnn}, demonstrate that GCNe, which incorporates edge features, achieves the best performance. Compared to the standalone attention network, the MAN+GCNe model shows significant improvements across all three datasets. Specifically, it enhances the F1-Scores by approximately 2.37\%, 2.00\%, and 1.28\%, while the B-Acc metric improves by about 2.06\%, 1.56\%, and 1.17\% respectively. Notably, the SAGEconv and GCN models exhibit a considerable performance gap compared to GCNe in joint training, even diminishing the effectiveness of MAN. This discrepancy may be attributed to these models' inability to effectively account for edge weights. On the other hand, GAT, which relies heavily on node connectivity, struggles to effectively infer node features in the context of Ethereum's transaction network, which is characterized by its sparsity.

\section{CONCLUSION}

In this paper, we introduced TLMG4Eth, a novel approach that integrates transaction language models with graph-based methods to capture semantic, similarity, and structural features of transaction data in Ethereum. Our work represents the first attempt to utilize language models to address the issue of unclear transaction semantics, and we pioneered the modeling of transaction similarity. After using an attention network to fuse semantic and similarity information, we proposed the construction of an account interaction graph to capture the structural information of the account transaction network. Furthermore, we developed a method for jointly training the attention network and the account structure graph to integrate information from all stages. Our approach has demonstrated significant improvements, achieving 10\% to 20\% performance gains across three datasets compared to current state-of-the-art methods. These empirical results provide strong evidence for the effectiveness of our proposed methodology, highlighting the potential of combining linguistic, semantic, and structural analysis in blockchain analytics and fraud detection.


\begin{thebibliography}{29}
\providecommand{\natexlab}[1]{#1}

\bibitem[{Ahmed et~al.(2019)Ahmed, Rossi, Lee, Willke, Zhou, Kong, and Eldardiry}]{role2vec}
Ahmed, N.~K.; Rossi, R.~A.; Lee, J.~B.; Willke, T.~L.; Zhou, R.; Kong, X.; and Eldardiry, H. 2019.
\newblock role2vec: Role-based network embeddings.
\newblock \emph{Proc. DLG KDD}, 1--7.

\bibitem[{{chainalysis}(2023)}]{chainalysis_report}
{chainalysis}. 2023.
\newblock 2023 Crypto Crime Trends: Illicit Cryptocurrency Volumes Reach All-Time Highs Amid Surge in Sanctions Designations and Hacking.
\newblock \url{https://www.chainalysis.com/blog/2023-crypto-crime-report-introduction/}.
\newblock Accessed: 2023-12-1.

\bibitem[{Chen et~al.(2019)Chen, Peng, Liu, Li, Xie, and Zheng}]{xblock}
Chen, L.; Peng, J.; Liu, Y.; Li, J.; Xie, F.; and Zheng, Z. 2019.
\newblock {XBLOCK Blockchain Datasets}: {InPlusLab} Ethereum Phishing Detection Datasets.
\newblock \url{http://xblock.pro/ethereum/}.

\bibitem[{Devlin(2018)}]{bert}
Devlin, J. 2018.
\newblock Bert: Pre-training of deep bidirectional transformers for language understanding.
\newblock \emph{arXiv preprint arXiv:1810.04805}.

\bibitem[{Grover and Leskovec(2016)}]{node2vec}
Grover, A.; and Leskovec, J. 2016.
\newblock node2vec: Scalable feature learning for networks.
\newblock In \emph{Proceedings of the 22nd ACM SIGKDD international conference on Knowledge discovery and data mining}, 855--864.

\bibitem[{Hamilton, Ying, and Leskovec(2017)}]{sage}
Hamilton, W.; Ying, Z.; and Leskovec, J. 2017.
\newblock Inductive representation learning on large graphs.
\newblock \emph{Advances in neural information processing systems}, 30.

\bibitem[{Hu et~al.(2024)Hu, Huang, Chow, Wei, Wu, and Liu}]{zipzap}
Hu, S.; Huang, T.; Chow, K.-H.; Wei, W.; Wu, Y.; and Liu, L. 2024.
\newblock ZipZap: Efficient Training of Language Models for Large-Scale Fraud Detection on Blockchain.
\newblock In \emph{Proceedings of the ACM on Web Conference 2024}, WWW '24, 2807–2816. New York, NY, USA: Association for Computing Machinery.
\newblock ISBN 9798400701719.

\bibitem[{Hu et~al.(2023)Hu, Zhang, Luo, Lu, He, and Liu}]{bert4eth}
Hu, S.; Zhang, Z.; Luo, B.; Lu, S.; He, B.; and Liu, L. 2023.
\newblock Bert4eth: A pre-trained transformer for ethereum fraud detection.
\newblock In \emph{Proceedings of the ACM Web Conference 2023}, 2189--2197.

\bibitem[{Kipf and Welling(2016)}]{GNN_r1}
Kipf, T.~N.; and Welling, M. 2016.
\newblock Semi-supervised classification with graph convolutional networks.
\newblock \emph{arXiv preprint arXiv:1609.02907}.

\bibitem[{Li et~al.(2022{\natexlab{a}})Li, Xie, Xu, Zhou, and Xuan}]{GNN_r4}
Li, P.; Xie, Y.; Xu, X.; Zhou, J.; and Xuan, Q. 2022{\natexlab{a}}.
\newblock Phishing fraud detection on ethereum using graph neural network.
\newblock In \emph{International Conference on Blockchain and Trustworthy Systems}, 362--375. Springer.

\bibitem[{Li et~al.(2022{\natexlab{b}})Li, Gou, Liu, Hou, Li, and Xiong}]{ttagn}
Li, S.; Gou, G.; Liu, C.; Hou, C.; Li, Z.; and Xiong, G. 2022{\natexlab{b}}.
\newblock TTAGN: Temporal transaction aggregation graph network for ethereum phishing scams detection.
\newblock In \emph{Proceedings of the ACM Web Conference 2022}, 661--669.

\bibitem[{Li et~al.(2023)Li, Gou, Liu, Xiong, Li, Xiao, and Xing}]{GNN_r3}
Li, S.; Gou, G.; Liu, C.; Xiong, G.; Li, Z.; Xiao, J.; and Xing, X. 2023.
\newblock TGC: Transaction Graph Contrast Network for Ethereum Phishing Scam Detection.
\newblock In \emph{Proceedings of the 39th Annual Computer Security Applications Conference}, 352--365.

\bibitem[{Lin et~al.(2021)Lin, Meng, Sun, Han, Kuang, Li, and Wu}]{bertgcn}
Lin, Y.; Meng, Y.; Sun, X.; Han, Q.; Kuang, K.; Li, J.; and Wu, F. 2021.
\newblock Bertgcn: Transductive text classification by combining gcn and bert.
\newblock \emph{arXiv preprint arXiv:2105.05727}.

\bibitem[{Liu et~al.(2024)Liu, Wang, Xu, Liu, Sun, Guo, and Ma}]{liu2024source}
Liu, R.; Wang, Y.; Xu, H.; Liu, B.; Sun, J.; Guo, Z.; and Ma, W. 2024.
\newblock Source Code Vulnerability Detection: Combining Code Language Models and Code Property Graphs.
\newblock \emph{arXiv preprint arXiv:2404.14719}.

\bibitem[{Lu, Du, and Nie(2020)}]{vgcn_bert}
Lu, Z.; Du, P.; and Nie, J.-Y. 2020.
\newblock VGCN-BERT: augmenting BERT with graph embedding for text classification.
\newblock In \emph{Advances in Information Retrieval: 42nd European Conference on IR Research, ECIR 2020, Lisbon, Portugal, April 14--17, 2020, Proceedings, Part I 42}, 369--382. Springer.

\bibitem[{Perozzi, Al-Rfou, and Skiena(2014)}]{deepwalk}
Perozzi, B.; Al-Rfou, R.; and Skiena, S. 2014.
\newblock Deepwalk: Online learning of social representations.
\newblock In \emph{Proceedings of the 20th ACM SIGKDD international conference on Knowledge discovery and data mining}, 701--710.

\bibitem[{Ramos et~al.(2003)}]{tfidf}
Ramos, J.; et~al. 2003.
\newblock Using tf-idf to determine word relevance in document queries.
\newblock In \emph{Proceedings of the first instructional conference on machine learning}, volume 242, 29--48. Citeseer.

\bibitem[{Shannon(1948)}]{npmi}
Shannon, C.~E. 1948.
\newblock A mathematical theory of communication.
\newblock \emph{The Bell system technical journal}, 27(3): 379--423.

\bibitem[{Tan et~al.(2021)Tan, Tan, Zhang, and Li}]{2vec_r1}
Tan, R.; Tan, Q.; Zhang, P.; and Li, Z. 2021.
\newblock Graph Neural Network for Ethereum Fraud Detection.
\newblock In \emph{2021 IEEE International Conference on Big Knowledge (ICBK)}, 78--85.

\bibitem[{Vaswani(2017)}]{attention}
Vaswani, A. 2017.
\newblock Attention is all you need.
\newblock \emph{arXiv preprint arXiv:1706.03762}.

\bibitem[{Veli{\v{c}}kovi{\'c} et~al.(2017)Veli{\v{c}}kovi{\'c}, Cucurull, Casanova, Romero, Lio, and Bengio}]{gat}
Veli{\v{c}}kovi{\'c}, P.; Cucurull, G.; Casanova, A.; Romero, A.; Lio, P.; and Bengio, Y. 2017.
\newblock Graph attention networks.
\newblock \emph{arXiv preprint arXiv:1710.10903}.

\bibitem[{Wang et~al.(2022)Wang, Chen, Xu, Wu, Shen, Xuan, and Yang}]{tsgn}
Wang, J.; Chen, P.; Xu, X.; Wu, J.; Shen, M.; Xuan, Q.; and Yang, X. 2022.
\newblock Tsgn: Transaction subgraph networks assisting phishing detection in ethereum.
\newblock \emph{arXiv preprint arXiv:2208.12938}.

\bibitem[{Wood et~al.(2014)}]{intro_r1}
Wood, G.; et~al. 2014.
\newblock Ethereum: A secure decentralised generalised transaction ledger.
\newblock \emph{Ethereum project yellow paper}, 151(2014): 1--32.

\bibitem[{Wu et~al.(2020)Wu, Yuan, Lin, You, Chen, Chen, and Zheng}]{2vec_r3}
Wu, J.; Yuan, Q.; Lin, D.; You, W.; Chen, W.; Chen, C.; and Zheng, Z. 2020.
\newblock Who are the phishers? phishing scam detection on ethereum via network embedding.
\newblock \emph{IEEE Transactions on Systems, Man, and Cybernetics: Systems}, 52(2): 1156--1166.

\bibitem[{Yuan et~al.(2020)Yuan, Huang, Zhang, Wu, Zhang, and Zhang}]{2vec_r2}
Yuan, Q.; Huang, B.; Zhang, J.; Wu, J.; Zhang, H.; and Zhang, X. 2020.
\newblock Detecting Phishing Scams on Ethereum Based on Transaction Records.
\newblock In \emph{2020 IEEE International Symposium on Circuits and Systems (ISCAS)}, 1--5.

\bibitem[{Zhang, Chen, and Lu(2021)}]{GNN_r5}
Zhang, D.; Chen, J.; and Lu, X. 2021.
\newblock Blockchain phishing scam detection via multi-channel graph classification.
\newblock In \emph{Blockchain and Trustworthy Systems: Third International Conference, BlockSys 2021, Guangzhou, China, August 5--6, 2021, Revised Selected Papers 3}, 241--256. Springer.

\bibitem[{Zheng et~al.(2020)Zheng, Zheng, Wu, and Dai}]{xblock_1}
Zheng, P.; Zheng, Z.; Wu, J.; and Dai, H.-N. 2020.
\newblock Xblock-eth: Extracting and exploring blockchain data from ethereum.
\newblock \emph{IEEE Open Journal of the Computer Society}, 1: 95--106.

\bibitem[{Zheng et~al.(2024)Zheng, Su, Chen, Lo, Zhong, and Ye}]{intro_r3}
Zheng, Z.; Su, J.; Chen, J.; Lo, D.; Zhong, Z.; and Ye, M. 2024.
\newblock Dappscan: building large-scale datasets for smart contract weaknesses in dapp projects.
\newblock \emph{IEEE Transactions on Software Engineering}.

\bibitem[{Zheng et~al.(2018)Zheng, Xie, Dai, Chen, and Wang}]{intro_r2}
Zheng, Z.; Xie, S.; Dai, H.-N.; Chen, X.; and Wang, H. 2018.
\newblock Blockchain challenges and opportunities: A survey.
\newblock \emph{International journal of web and grid services}, 14(4): 352--375.

\end{thebibliography}

\clearpage

\end{document}